\documentclass{cimento}

\usepackage{graphicx}

\title{From neutron star binaries to gamma-ray bursts}

\author{S.Rosswog\from{ins:x}}
\instlist{\inst{ins:x} School of Engineering and Science, International University Bremen, Germany}

\usepackage{epsfig}

\begin{document}

\def\Mesz{M\'esz\'aros~}
\def\Pacz{Paczy\'nski~}
\def\Kluz{Klu\'zniak~}
\def\p{$e^\pm \;$}
\def\msun{M$_{\odot}$}

\maketitle

\begin{abstract}
I summarize recent results about how a neutron star binary coalescence can
produce short gamma-ray bursts (GRBs). Two possibilities are discussed: the
$\nu_i \bar{\nu}_i$-annihilation above the merged remnant and the exponential
amplification of magnetic fields in the central object up to values close to
equipartition. \\
We find that the annihilation of $\nu_i \bar{\nu}_i$-pairs drives bipolar,
relativistic outflows with Lorentz-factors large enough to circumvent the
GRB 'compactness problem'. The total energy within these outflows is moderate
by GRB-standards ($\sim 10^{48}-10^{49}$ ergs), but the interaction with
the baryonic material blown-off by the neutrinos collimates the outflows into opening angles of
typically $0.1$ sterad, yielding isotropic energies close to $10^{51}$ 
ergs.\\ 
We further want to stress the plausibility of the central object resisting the
immediate collapse to a black hole. In this case the central object will
--similar to a 
proto-neutron star-- be subject to neutrino driven convection that --together
with the rapid, differential rotation-- will lead to a drastic amplification of
pre-existing magnetic fields. Within fractions of a second, field strengths
comparable to equipartition field strength ($> 10^{17}$ G) will be
reached. These will produce large torques that will spin-down the object
within about 0.2 s, and would thus naturally explain the duration of short GRBs.
\end{abstract}

\section{Introduction}

Since nearly two decades coalescences of double neutron star systems (DNSs)
are thought to cause gamma-ray bursts (GRBs). This possibility has been
mentioned by \Pacz (1986), Goodman (1986), Goodman et al. (1987) and discussed
in  detail by Eichler et al. (1989) and Narayan et al. (1992) (for a more
detailed bibliography see the reviews of \Mesz (2002) and Piran (2005)).\\
The coalescence of two neutron stars releases a gravitational binding energy of
a few times $10^{53}$ ergs and therefore more than enough to power a short GRB
with a gamma-ray energy of $E_{\gamma} \sim 10^{51} \left( \frac{\Omega}{4
    \pi}\right)$ ergs, $\Omega$ being the solid angle into which the photons are
emitted. Due to its compactness such a merger can produce substantial
variations in  its energy output on very short time scales: a neutron-star
like object has a dynamical time scale $\tau_{\rm dyn, ns}= (G
\bar{\rho})^{-1/2}= 0.4 \; (\bar{\rho}_{14})^{-1/2}$ ms, the orbital time
scale at the marginally bound orbit of a new-born black hole is $\tau_{\rm
  dyn, bh} \approx \frac{16 \pi G  M_{\rm BH}}{c^3} = 0.2 \left( \frac{M_{\rm
      BH}}{2.5 {\rm M}_\odot} \right)$ ms. \\
For several years there has been a discrepancy of about
one order of magnitude between rates estimated based on the number of observed
systems and those derived from population synthesis models. The recent
discoveries of several new DNS (Stairs 2004, Faulkner 2005) have substantially
increased the observation-based 
estimates, now being in good agreement with the population synthesis-estimate
of about $\sim 10^{-4}$ events per year and galaxy (Kalogera et al. 2004).
 The rate of GRBs is estimated to be $\sim 10^{-7}$ per year and galaxy
with about 25 \% of which belong to the short variety of bursts (Hurley et
al. 2002). Therefore, even if most DNS mergers should fail to produce
detectable GRBs (either due to very narrow beaming or failure to produce a GRB
at all), the neutron star merger rate would still be high enough to explain
all short GRBs. \\
Most often it is assumed that the remnant of a DNS coalescence will
immediately form a black hole that is surrounded by an accretion disk of
neutron star debris. Therefore the coalescence of a neutron star black hole
system, to our knowledge first suggested by \Pacz (1991), is generally only
considered to be a variation on the DNS merger theme. 
Although this is a reasonable assumption, we want to point out two problems
with this scenario. First, there is a controversy about the rate of such
events. Bethe and Brown (1998) estimated rates as high as $10^{-4}$ per year
and galaxy while a recent study (Pfahl et al. 2005) estimates that there is
less than one black hole pulsar system per 100 DNS systems. This fraction
would be consistent with the non-observation of black hole 
neutron star systems while there are to date 8 DNS systems known. Second, even
if their number should in principle be sufficient, the accretion process is
considerably 
more complex than in the DNS case as there is a much larger range of possible
mass ratios, leading to mass transfer whose dynamics is governed by the
interplay between gravitational wave emission backreaction (trying to decrease
the orbital separation), mass transfer (trying to widen the orbit) and the
reaction of the neutron star to mass loss (depending on the nuclear equation
of state and on the current neutron star mass). Using conventional nuclear
physics, i.e. neutrons and protons as the only hadronic constituents of
neutron star matter, we found it difficult (Rosswog et al. 2004, Rosswog 2005)
to produce accretion disks that are likely to launch a GRB. This issue will
have to be explored further in the future.   

\section{Pathways to a GRB}
\begin{figure}
\centerline{\psfig{file=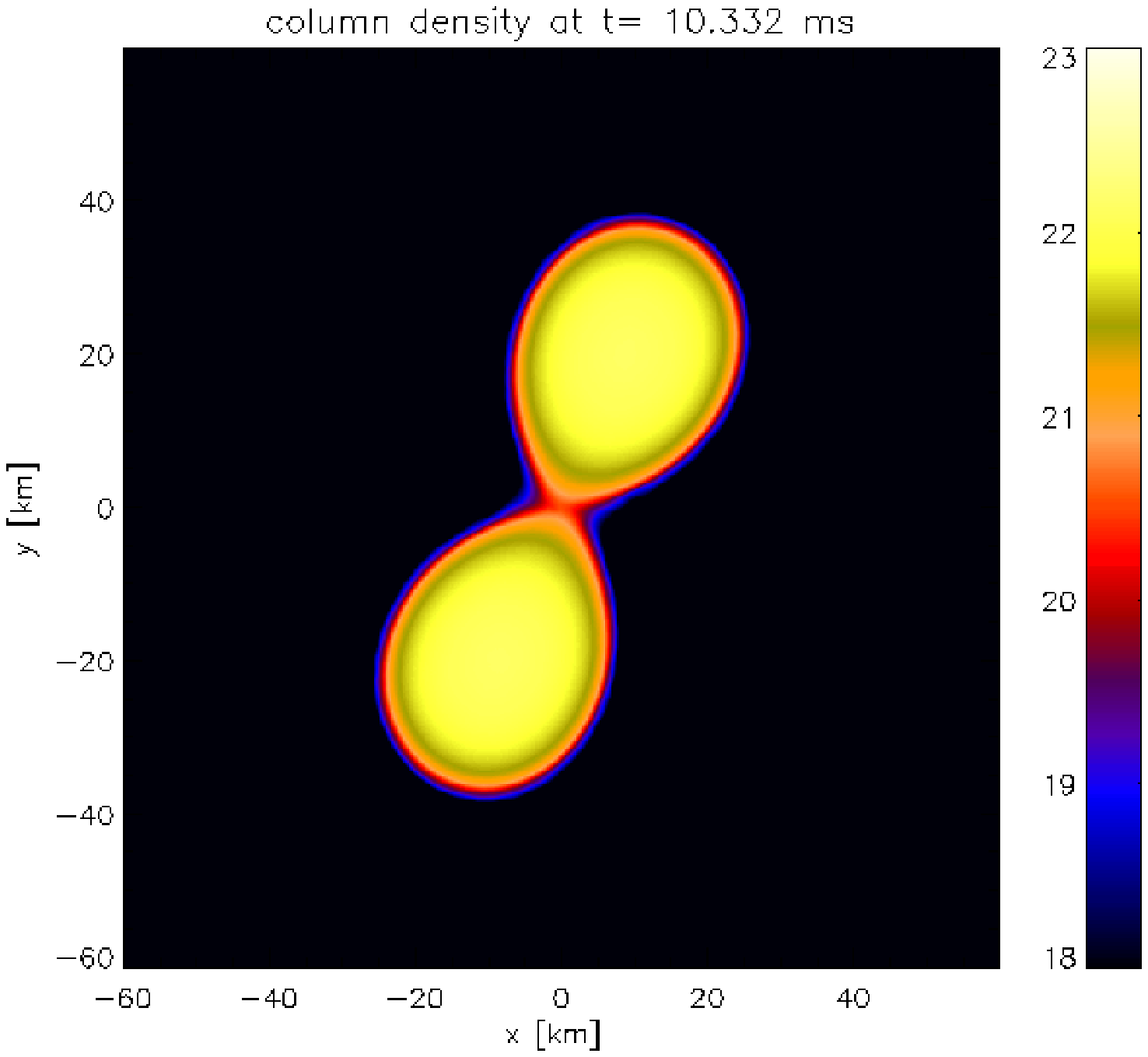,width=6cm,angle=0}
           \psfig{file=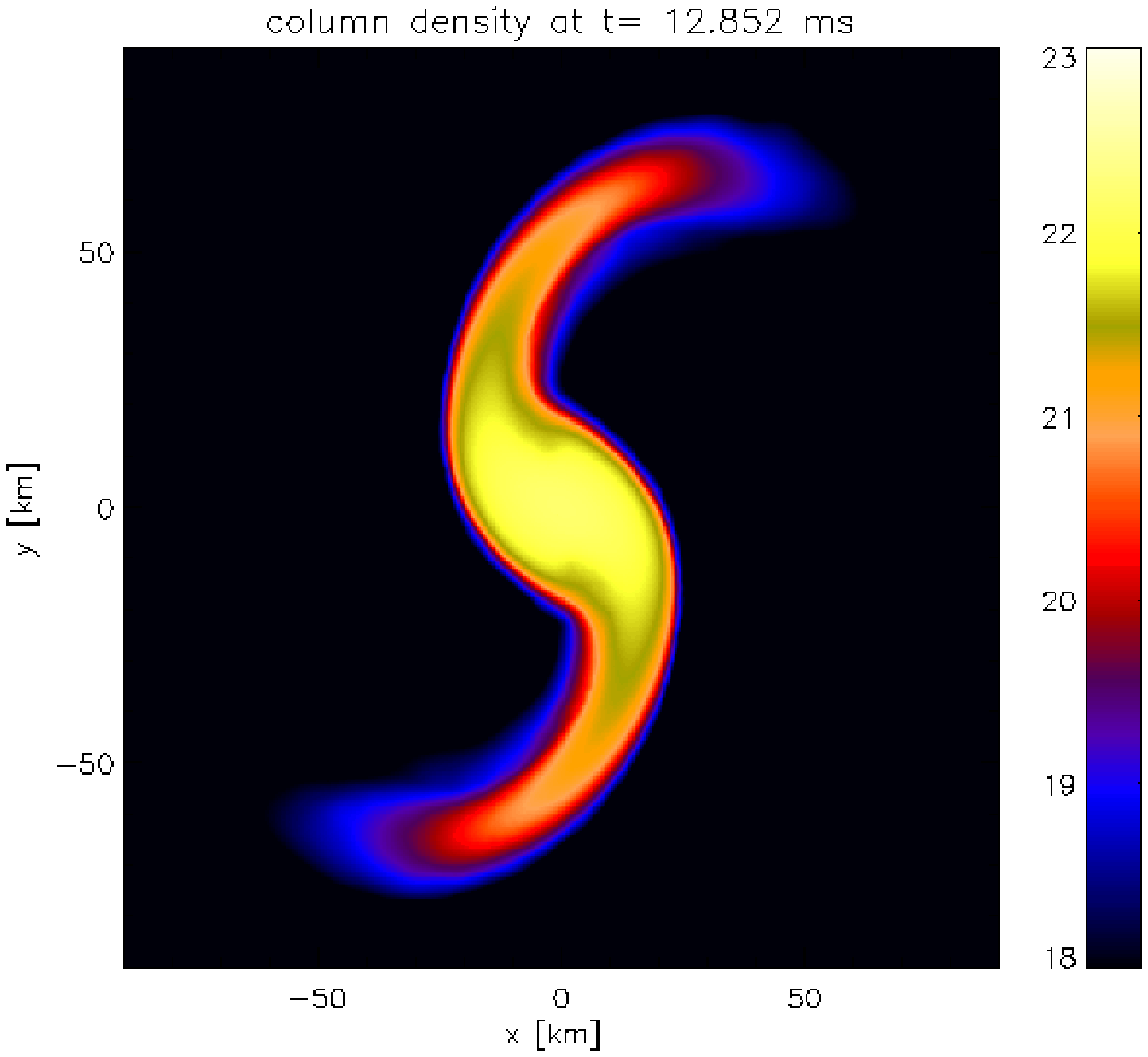,width=6cm,angle=0}
           \psfig{file=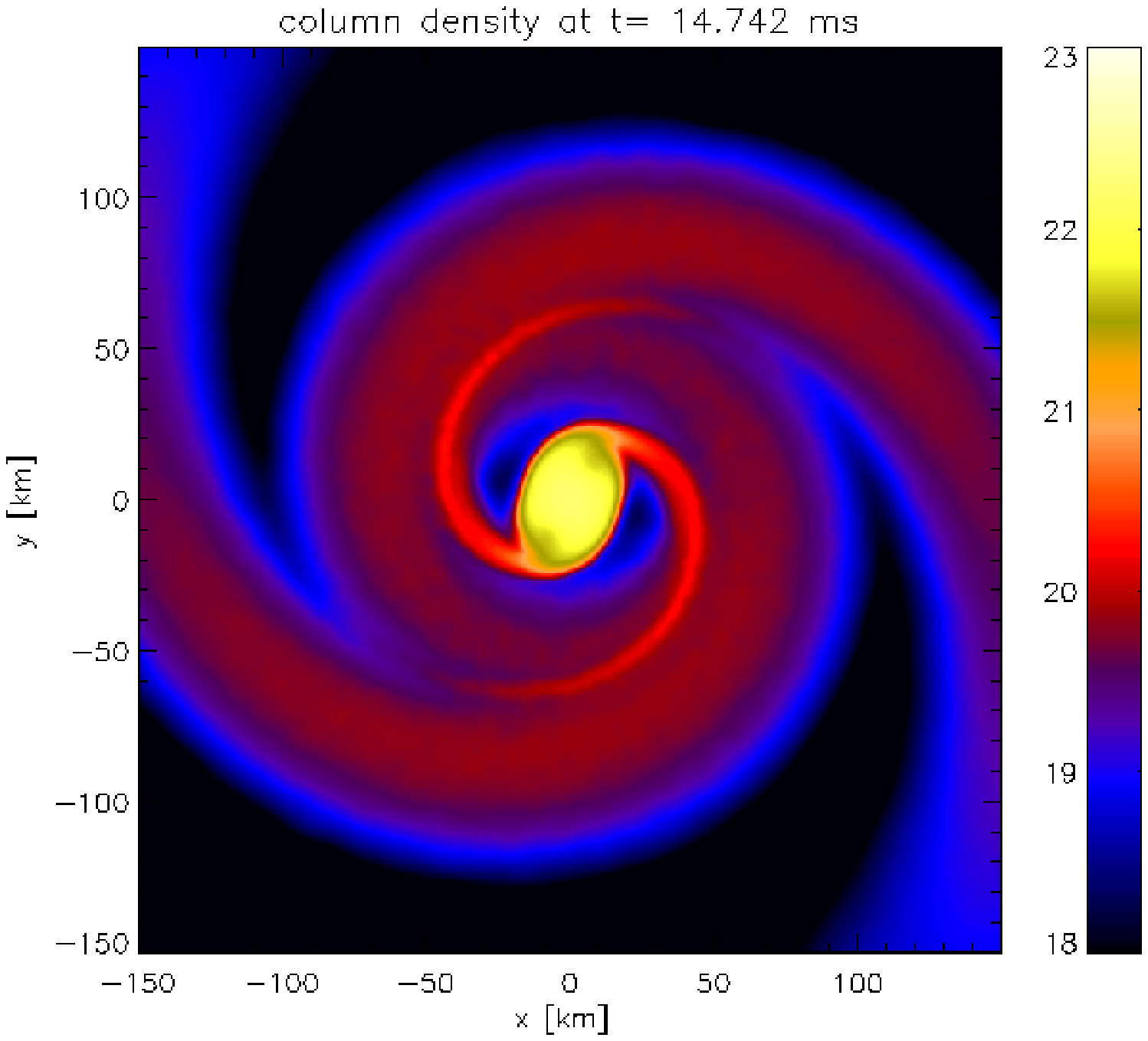,width=6cm,angle=0}}
\caption{Coalescence of binary neutron star system (both stars 1.4
           M$_{\odot}$, tidal locking as initial spin state). Colour-coded is
           column density.\label{coldens}}
\end{figure}
We will now address possible pathways that lead from the coalescence, see Fig.
1, to the production of the ultra-relativistic outflow that is required to
accommodate the large energies and the variations on millisecond
time scales with non-thermal spectra. To be consistent with
observations, the GRBs are required to have Lorentz-factors $\Gamma$ of
several hundreds (Litwick and Sari 2001). Such Lorentz-factors can only be
obtained if the available energy, $E$, is deposited in a volume that contains
only a small amount of
baryonic material, $m \sim E / \Gamma c^2$. Possible mediators to transport
the energy into a baryon-poor region are the neutrinos that are
produced in the hot neutron star debris and/or magnetic fields.\\
We will discuss here two possibilities: the neutrino-annihilation above the
merger remnant as found in our simulations (supermassive neutron star plus
disk) and a possible dramatic increase in the magnetic fields strength within
the temporarily stabilized central object produced in the merger.\\
The calculations we refer to use a 3D smoothed particle hydrodynamics (SPH)
with Newtonian self-gravity and gravitational wave back reaction forces. 
We use a temperature and composition dependent EOS and account for changes in
the electron fraction and cooling due to neutrinos by a detailed multi-flavour
neutrino scheme. Particular attention has been payed to the implementation of artificial
viscosity: it is only present where needed, i.e. in shocks, and absent
otherwise. For a detailed description of the numerical techniques we refer to
Rosswog et al. (2000), Rosswog and Davies (2002), Rosswog and Liebend\"orfer
(2003) and Rosswog et al. (2003).

\subsection{On the beaten track: annihilation of $\nu_i \bar{\nu}_i$-pairs}
The remnants of a DNS merger radiate neutrinos at a total luminosity of $\sim
2 \cdot 10^{53}$ ergs/s. The neutrino energies are between 10 and 25 MeV, the
luminosities are clearly dominated by electron anti-neutrinos (Rosswog and
Liebend\"orfer 2003). Similar results, but somewhat higher luminosities have
been found by Ruffert and Janka (2001). Neutrino and anti-neutrinos annihilate
above the merger remnant, yielding roughly elliptical contours of deposited
energy per time and volume (see Figure 2, right column in Rosswog et
al. 2003). It is the energy deposition per rest mass, $\eta$, that
determines the asymptotic Lorentz-factor of the outflow. Guided by the typical
duration of short GRBs we have assumed a deposition time, $\tau_{\rm dep}= 1$ 
s, to estimate $\eta= \frac{Q_{\nu\bar{\nu}}\tau_{\rm dep}}{\rho c^2}$, where
$Q_{\nu\bar{\nu}}$ is the annihilation energy per time and
volume. This is shown in Figure 2 for the case of the neutron star
binary system whose evolution is shown in Figure 1.  We find peak asymptotic Lorentz-factors of more than
$10^4$. This particular, initially corotating system is probably on the
optimistic side as its large angular momentum yields a particularly broad
funnel region above the central object. Nevertheless it shows that the
neutrino annihilation mechanism has no problems in producing outflows with
the required Lorentz-factors of several hundreds.
\begin{figure}
\centerline{\psfig{file=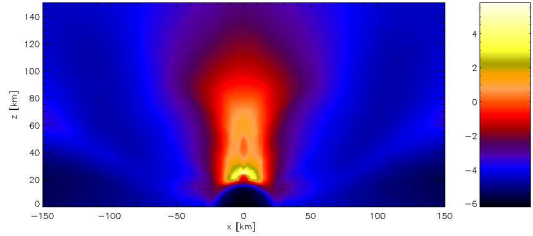,width=6cm,angle=90}}
\caption{Colour-coded is the logarithm of the energy deposition per rest mass,
  $\log_{10}(\eta)$, from $\nu_i \bar{\nu}_i$-annihilation for the simulation shown in the previous figure. The quantity $\eta$ determines the asymptotic
  Lorentz-factor, $\Gamma_{\rm asym}\approx \eta$.\label{gamma}}
\end{figure}\\
The total energy contained in the outflow, however, is
substantially lower, $\sim10^{48} -  10^{49}$ ergs, than the estimated isotropic energy
equivalent of $\sim 10^{51}$ ergs, that is estimated for short GRBs at a 
redshift of about unity (e.g. Panaitescu et al. 2001). Therefore, the outflow
has to be well-collimated in 
order to explain the above mentioned isotropic energy equivalent. Although
well-collimated outflows occur in a large variety of astrophysical
environments there is no generally accepted mechanism how this is achieved. 
At luminosities in excess of $10^{53}$ ergs/s the neutrinos emitted from the
remnant will drive an energetic baryonic wind that will engulf the remnant.
We have investigated to which extent this material can collimate the outflow
(Rosswog and Ramirez-Ruiz, 2003). We find that the result is quite sensitive
to the total mass of the binary system, which governs the total neutrino
luminosity. As the baryonic wind is sensitive to the neutrino luminosity, even
a rather narrow distribution of DNS masses will 
result in a broad distribution of opening angles. We find typical opening
angles of $\sim 0.1$ sterad, transforming into typical isotropized
energies close to $10^{51}$ ergs (see Figure 3 in Rosswog and
Ramirez-Ruiz, 2003).\\
Recently, Aloy et al. (2004) have performed relativistic simulations to infer
the launch of jets from DNS merger remnant disks. Their detailed simulations
yield results in close agreement with our above estimates.\\
In summary, neutrino annihilation from DNS remnants seems to be able to launch
relativistic jets with Lorentz-factors in excess of 100. The resulting
outflows are relatively weak by GRB-standards, but with typical opening angles
of about 0.1 sterad they yield isotropized energies in close agreement with
the $10^{51}$ ergs that are inferred from observations. 

\subsection{Off the beaten track: ``hyper-magnetars''}
In this section we want to discuss the possibility of a GRB being launched by
a meta-stable, hyper-magnetized neutron star-like object. Further ideas about
GRBs launched from strongly magnetized compact objects can be found in the
literature (e.g. Usov 1992, 1994, Duncan and Thompson 1992, Thompson and 
Duncan 1994, \Mesz and Rees 1997, Katz 1997, Kluzniak and Ruderman
1998 and Lyutikov and Blandford 2003).\\ 
The merger remnant harbors a central object of about 2.5 \msun, a mass that
is beyond the maximum mass of most physical equations of state. Therefore, the
central object will probably collapse at some stage to a black hole. Our
hydrodynamic simulations (Rosswog and Davies 2002) show that the central
object is rotating differentially, with periods from 0.3 ms to about 2ms from
the center to the edge of the central object. As differential rotation can
yield centrifugal support without mass shedding, it is very efficient in
stabilizing configurations beyond their non-rotating upper mass limit.\\
The lifetime of the metastable central object depends on the time scale
necessary to damp out the differential rotation. A discussion of these time
scales and references
can be found in  Rosswog and Davies (2002). Although this question is far
from being settled, we consider it plausible that the remnant remains stable
for about a second, i.e. long enough for a short GRB. If so, this
will open up possibilities to launch GRBs from the extreme, but probably
short-lived magnetic fields in the central object (similar effects with
down-scaled field strengths may still occur in the remnant disk, independent of
whether the central object remains stable or not; for a comparison of the
maximum field strengths across the merger remnant see Fig. 8 in Rosswog et
al. 2003).\\
Like in a proto-neutron star born in a supernova explosion, the neutrino
emission from the remnant will build up entropy and lepton number gradients
that will drive convective motion. If the rotation periods, $\tau_{\rm rot}$,
are smaller than the convective overturn times, $\tau_{\rm conv}$, the
remnant will support dynamo action. From our simulations we found a ratio,
$Ro= \tau_{\rm rot}/\tau_{\rm conv}$ well below unity and therefore expect a
dynamo to be active in the remnant. The field strength is expected to grow
exponentially with an e-folding time close to the convective overturn time
scale until it becomes buoyant. If initial neutron star field strengths of
$10^{12}$ G are assumed, equipartition field strength ($> 10^{17}$ G) will be
reached after only 40 ms, the field will then become buoyant and
float up. If the whole rotational energy ($E_{\rm rot}\sim 8 \cdot 10^{52}$
erg) is transformed into magnetic field 
energy, the field strength, averaged over the central object, will be
$\bar{B}_{co} \sim 3\cdot 10^{17}$ G (still $10^{17}$ G, if only 10 \% are transformed into
magnetic energy). Owing to the hydrodynamic flow pattern within the central
object, see Fig. 8 in Rosswog and Davies (2002), we consider it most likely
that the field will be wound up in local vortices up to equipartition field
strengths and then will float up. Once it breaks through the remnant surface
it will cause a sequence of relativistic blasts, the time structure of which
will be governed by the fluid instabilities.\\
To estimate the time scale on which the central object will be spun down,
$\tau_{sd}$, we use the averaged field strength $\bar{B}_{co}$, the
geometry found in the simulations and the magnetic dipole luminosity, $L_{\rm
  dp}$. We find $\tau_{sd}= E_{\rm rot}/L_{\rm dp}\approx 0.2$ s, 
very close to the typical duration of a short GRB.

\section{Summary}
We have discussed possible pathways leading from the coalescence of a
double neutron star system to the ultra-relativistic outflows that will later,
far away from the central source, produce a (short-variety) GRB. In
particular, we have addressed two mechanisms: the annihilation of $\nu_i
\bar{\nu}_i$-pairs in the centrifugally cleaned pole regions above the merger
remnants and the amplification of the initial neutron star magnetic fields via
a convective dynamo inside the temporarily stabilized central object of the
remnant.\\
We find that the $\nu_i \bar{\nu}_i$-annihilation will drive bipolar,
relativistic jets with large Lorentz-factors (in extreme cases beyond
$10^{4}$). The total energies within these jets are moderate by GRB-standards,
$\sim 10^{48} ... 10^{49}$ ergs, but due to the interaction with baryonic material that
has been blown off the remnant via the intense neutrino radiation, these
outflows are typically collimated into half-opening angles of about 0.1 sterad.
An observer would infer isotropized luminosities of about $10^{51}$ ergs.\\
We consider it to be very plausible that the central object in the remnant
is stabilized by differential rotation for long enough to launch a GRB before
collapsing to a BH. If it remains stable for a good fraction of a second then the initial neutron
star magnetic fields are expected to be amplified by a low-Rossby number  $\alpha-\Omega$-dynamo. 
In principle, enough rotational energy is available 
to attain an average field strength in the central object of $3 \cdot 10^{17}$ G. Locally 
the equipartition field strength (ranging from $10^{16}$ to a few times $10^{17}$ G depending 
on the exact position in the remnant) may be reached. This will cause the corresponding 
fluid parcels to float up and produce via reconnection an erratic sequence of ultra-relativistic 
blasts. In addition, the central object can act as a ``super-pulsar'' of $\sim 10^{17}$ G
that transforms most of its rotational energy into an ultra-relativistic wind with frozen-in 
magnetic field. As shown in Usov (1994) such a wind will result in a black-body component
plus synchro-Compton radiation. Such a super-pulsar will spin-down in $\sim 0.2$ s, just the
typical duration of a short GRB.

\end{document}